\documentstyle[ltwol]{article}

\def\Journal#1#2#3#4{{#1} {\bf #2}, #3 (#4)}

\def\PLB{{\em Phys. Lett.}  B}
\def\PRL{\em Phys. Rev. Lett.}
\def\PRD{{\em Phys. Rev.} D}

\def\be{\begin{equation}}
\def\ee{\end{equation}}
\def\bea{\begin{eqnarray}}
\def\eea{\end{eqnarray}}

\begin{document}

\title{SCALAR-MEDIATED FLAVOR-CHANGING NEUTRAL CURRENTS}

\author{MARC SHER}

\address{Physics Dept., College of William and Mary, Williamsburg VA 23187 USA
\\E-mail: sher@physics.wm.edu}


\twocolumn[\maketitle\abstracts{ The simplest extension of the standard model
involves adding a scalar doublet--the so-called two-Higgs model.  In general,
the additional scalar will mediate tree-level flavor-changing neutral
currents.  Although one can arbitrarily impose a discrete symmetry to avoid
these, it isn't necessary to do so; reasonable assumptions about the size of
the flavor-changing couplings can make them sufficiently small as to avoid
problems in the kaon sector.  However, these same assumptions give much larger
effects in the third-family case.  We discuss the model, the ``reasonable
assumptions" on the size of the couplings, and examine phenomenological bounds on
the couplings, showing that the most promising signatures are from
$B_s\rightarrow
\mu\tau$, $\mu\rightarrow e\gamma$, etc.  We then include the newest result which
shows potentially significant effects on the anomalous magnetic moment of the
muon.}]

One of the simplest, and most-often studied, extensions of the standard model
is the ``two-Higgs doublet" model, in which an additional scalar doublet is
included.  A potentially serious problem with such models is the existence of
tree-level flavor-changing neutral currents.  Although one can eliminate such
currents with a discrete symmetry, such a symmetry is ad hoc, and may have
cosmological problems.  In this talk, I'll point out that the problem of
flavor-changing neutral currents (FCNC) may not be as serious as often
believed, and that reasonable assumptions about the structure of Yukawa
coupling matrices allow sufficient suppression of FCNC  couplings between the
first two generations, while simultaneously predicting much bigger
effects in coupling the second and third generations.  The phenomenological
bounds will be reviewed and the most promising decays examined.  Finally, a new
bound, from the anomalous magnetic moment of the muon, will be discussed.

It is easy to see why FCNC do not appear in the standard model.  The Yukawa
interactions in the standard model are of the form (focusing on the neutral
fields only)
\begin{equation}
{\cal L}_Y=-{h_{ij}\over\sqrt{2}}\overline{\psi}_i\psi_j\phi
\end{equation}
After spontaneous symmetry breaking, this gives a mass matrix
\begin{equation}
M_{ij}={h_{ij}\over\sqrt{2}}\langle\phi\rangle
\end{equation}
Clearly, when the mass matrix is diagonalized, the Yukawa coupling matrix will
also be diagonalized, and thus there will be no neutral flavor-changing
couplings involving the Higgs boson.

In the two doublet model, this is not the case.  The Yukawa interactions are of
the form
\begin{equation}
{\cal L}_Y=-{f_{ij}\over\sqrt{2}}\overline{\psi}_i\psi_j\phi_1
-{g_{ij}\over\sqrt{2}}\overline{\psi}_i\psi_j\phi_2
\end{equation}
which gives a mass matrix
\begin{equation}
M_{ij}={f_{ij}\over\sqrt{2}}\langle\phi_1\rangle+
{g_{ij}\over\sqrt{2}}\langle\phi_2\rangle
\end{equation}
It is clear that diagonalizing $M_{ij}$ will {\bf not} generally diagonalize
$f_{ij}$ and $g_{ij}$, leading automatically to flavor-changing neutral
couplings.  These models will thus have a tree-level contribution to, for
example, $K^o-\overline{K}^o$ mixing.

There are three solutions to this problem that have been discussed.  The first
two make use of the Glashow-Weinberg theorem\cite{glashowweinberg}, which
states that neutral flavor-changing couplings will be absent if all of the
quarks of a given charge couple only to a single Higgs doublet.   In Model I,
one imposes a discrete symmetry to couple only one doublet to the fermions.  In
Model II, one imposes a discrete symmetry to couple the $Q=2/3$ quarks to one
doublet and the $Q=-1/3$ quarks to the other (in supersymmtry, this discrete
symmetry arises automatically).  These both eliminate tree-level FCNC. 
However, they are {\it ad hoc}, with no justification other than eliminating
FCNC, and discrete symmetries may have problems with cosmological domain walls.
The third option, Model III, will be the focus of this talk.  In this model, no
discrete symmtries are introduced, and one looks at the phenomenological
constraints on the flavor changing couplings.

To define the flavor-changing couplings, one first chooses a basis so that
$\langle\phi_2\rangle=0$.   In that case,
$M_{ij}=f_{ij}\langle\phi_1\rangle/\sqrt{2}$, and then diagonalization of
$M_{ij}$ will automatically diagonalize the couplings of $\phi_1$.  The
couplings of $\phi_2$ remain arbitrary:
\begin{eqnarray}
\xi_{ij}^U\overline{Q}_{i,L}\tilde{\Phi}_2U_{j,R}+{\rm h.c.}\cr
\xi_{ij}^D\overline{Q}_{i,L}{\Phi}_2D_{j,R}+{\rm h.c.}
\end{eqnarray}
with a similar term for leptons.  Note that the choice of basis will not, in
general, diagonalize the Higgs mass matrices, and thus one should replace
$\Phi_2$ with each mass eigenstate times an appropriate mixing angle.  Since
the lightest mass eigenstate will dominate in the phenomenological effects, and
since the couplings are arbitrary, this angle will be ignored.

The biggest danger from the above couplings is the contribution to
$K^o-\overline{K}^o$ mixing, due to tree-level $\phi_2$ exchange.  Prior to
1987, it was claimed that the observed value of the mixing would be
inconsistent with the presence of this effect unless the mass of the $\phi_2$
was greater than $100$ TeV.  Since that is far beyond the electroweak scale,
Model III was excluded.

However, as pointed out by Cheng and Sher\cite{chengsher}, this result was
based on the assumption that $\xi_{ds}^D$ was of the order of the gauge
coupling, which is certainly not reasonable for a Yukawa coupling involving the
lightest generations.  They noted that in a wide variety of models, if one
assumes that there is no fine-tuning (in which several large numbers add or
subtract to form a small number), then the flavor-changing couplings will be of
the order of the geometric mean of the Yukawa couplings of the two generations.
Thus, writing
\begin{equation}
\xi_{ij}=\lambda_{ij}{\sqrt{m_im_j}\over \langle\phi_2\rangle}
\end{equation}
they argued that the most natural value of the $\lambda_{ij}$ is unity.  This
ansatz then strongly suppresses the value of $\xi_{ds}$, weakening the lower
bound on $\phi_2$ into the hundreds of GeV range.

Since we do not yet have a theory of flavor, it would be premature to make any
firm assumptions about the magnitude of the $\lambda_{ij}$, rather, one should
look for experimental signatures and bounds on these parameters, keeping in
mind that in a variety of models, their value should be approximately of order
one.

The most extensive discussion of the bounds on the $\lambda_{ij}$ was given in
a recent paper by Atwood, Reina and Soni\cite{ars}; most of these results are
discussed in more detail there.   The first process to examine is
$F-\overline{F}$ mixing, where $F=(K,D,B)$.  For a pseudoscalar mass of $500$
GeV (the pseudoscalar gives the strongest bounds, since the matrix element is
largest), the bounds are $\lambda_{ds}<0.2$, $\lambda_{db}<0.25$ and
$\lambda_{uc}<0.6$.  One can see that the bounds involving the first generation
are {\it very} stringent.

What about processes involving the third generation?  Atwood, Reina and Soni
show that the most important are $B_s\rightarrow \mu^+\mu^-$ and $B\rightarrow
X_s\mu^+\mu^-$.  The former gives a bound $\lambda_{sb}\lambda_{\mu\mu}< 9$ and
the latter gives a bound $\lambda_{sb}\lambda_{\mu\mu}< 4$.  The former bound
depends on the pseudoscalar mass, and the latter on the scalar mass (a value of
$200$ GeV was chosen).  Thus, these processes are getting into the interesting
region.   It was shown by Aliev, et al.\cite{aliev} that the bound from
$B\rightarrow X_s\gamma$ is not useful unless $\lambda_{bb}$ is very large.

It should be pointed out that the above ansatz for the couplings indicate that
the $tc\phi_2$ coupling could be very large.  In papers by Bar-Shalom,et
al.\cite{barshalom}, the processes $e^+e^-\rightarrow
\overline{t}c\nu_e\overline{\nu}_e$ and $\mu^+\mu^-\rightarrow \overline{t}c$
are examined; Han, et al.\cite{han} examined the processes $t\rightarrow
(g,Z,\gamma)$, and Hou\cite{hou} looked at the unusual process
$e^+e^-\rightarrow tt\overline{c}\overline{c}$.  All of these processes, of
course, require substantial top quark production.

In all of the above analyses, it was assumed that the flavor-changing couplings
occurred in the quark sector.  There is no reason, of course, why they should
not also occur in the lepton sector.  As pointed out by Sher and
Yuan\cite{sheryuan}, standard unification arguments should give relationships
between, say, $\lambda_{sb}$ and $\lambda_{\mu\tau}$.  They examined these
relationships, and found bounds on some rather unusual (and therefore
experimentally distinct) processes.

For example, from $B_s\rightarrow\mu\tau$ and $B\rightarrow K\mu\tau$, they
found that $\sqrt{\lambda_{sb}\lambda_{\mu\tau}}< 10$; and from 
$B_s\rightarrow\tau\tau$ and $B\rightarrow K\tau\tau$, 
$\sqrt{\lambda_{sb}\lambda_{\tau\tau}}< 30$.  Although these bounds are
somewhat weaker than those involving just muons, they (especially
$B_s\rightarrow \mu\tau$) are very experimentally distinct, and can thus be
significantly improved at a B-factory.  In fact, the current bound on
$B_s\rightarrow \mu\tau$ was guessed from a published bound on $B\rightarrow
\mu\mu X$; as yet there are {\bf no published bounds} on the {\bf most}
distinct and interesting process!

In the purely leptonic sector, the strongest bound\cite{sheryuan} comes from 
$\mu\rightarrow e\gamma$, with a $\tau$ intermediate state.   This gives
$\sqrt{\lambda_{\mu\tau}\lambda_{e\tau}}<5$.  They did not calculate the
contribute to muon-electron conversion off a nucleus, which could give stronger
bounds.  Processes involving rare $\tau$ decays give extremely weak bounds.

Note that all of the bounds above depend on the product of two different
$\lambda$'s, with the exception of $F-\overline{F}$ mixing (which bounds
$\lambda$'s which have a first generation index).  Since one might expect the
biggest contribution to come from coupling between the second and third
generations ($\lambda_{sb},\lambda_{\mu\tau},\lambda_{ct}$), it is desirable to
look for processes which only depend on a single coupling.  Such a process was
recently discussed by Nie and Sher\cite{niesher}.  They noted that one can look
at $g-2$ for the muon.  If one looks at the vertex correction, where the muons
exchange a scalar and turn into $\tau$'s, then this contribution will only
depend on $\lambda_{\mu\tau}$.  Using current bounds, they find that
$\lambda_{\mu\tau}<50$; but this will be improved by a factor of $5$ in an
upcoming experiment.  One might think that $\lambda_{\mu\tau}<10$ is not an
impressive bound, but recall that the above ansatz is just a guess, and that
the actual bound on $\xi_{\mu\tau}\equiv \lambda_{\mu\tau}(\sqrt{m_\mu
m_\tau}/v)$ would be $\xi_{\mu\tau}<{1\over 40}$, which is certainly in the
interesting range.

In summary, one of the simplest extensions of the standard model is the
addition of a scalar doublet.  In general, this yields tree-level FCNC. 
Defining couplings as $\xi_{ij}\equiv \lambda_{ij}(\sqrt{m_im_j}/v)$, theorists
would generally expect $\lambda_{ij}$ to be of order 1, however in view of our
lack of understanding of flavor physics, one should look at the
phenomenological bounds on these FCNC processes.  Bounds from $F-\overline{F}$
mixing show that couplings involving the first generation are quite small. 
Couplings involving the second and third generations can be much larger.  The
strongest and most intriguing bound (because it has no one-loop counterpart in
the standard model) is from $B_s\rightarrow\mu\tau$, which is not
experimentally bounded in the literature.  This bound, and others, depends on
the product of two $\lambda$'s.  A recent analysis of the anomalous magnetic
moment of the muon showed that it is only sensitive to $\lambda_{\mu\tau}$, and
not to a product of two couplings.


\section*{References}
\begin{thebibliography}{99}
\bibitem{glashowweinberg} S.L. Glashow and S. Weinberg,
\Journal{\PRD}{15}{1958}{1977}.

\bibitem{chengsher}, T.P. Cheng and M. Sher, \Journal{\PRD}{35}{3484}{1987}.

\bibitem{ars} D. Atwood, L. Reina and A. Soni, \Journal{\PRD}{55}{3156}{1977}.

\bibitem{aliev} T.M. Aliev and E.O. Iltan, hep-ph/9803272.

\bibitem{barshalom} S. Bar-Shalom, G. Eilam and A. Soni,
\Journal{\PRD}{57}{2957}{1998}; \Journal{\PRL}{79}{1217}{1997}.

\bibitem{han} T. Han, K. Whisnant, B.L. Young and X. Zhang,
\Journal{\PLB}{385}{311}{1996}.

\bibitem{hou} W.-S. Hou, G.-L. Lin and C.-Y. Ma, \Journal{\PRD}{56}{7434}{1997}.

\bibitem{sheryuan}  M. Sher and Y. Yuan, \Journal{\PRD}{44}{1461}{1991}.

\bibitem{niesher} S. Nie and M. Sher, \Journal{\PRD}{58}{11xxxx}{1998}.

\end{thebibliography}
\end{document}